\documentclass[%
 preprint,
 amsmath,amssymb,
 aps,prl
]{revtex4-1}

\usepackage{graphicx}% Include figure files
\usepackage{dcolumn}% Align table columns on decimal point
\usepackage{bm}% bold math
\begin{document}

\preprint{APS/123-QED}

\title{Experimental verification of the rotational sense and type of chiral spin spiral structure by spin-polarized scanning tunneling microscopy}% Force line breaks with \\

\author{Masahiro Haze}

\author{Yasuo Yoshida}%
 \email{yyoshida@issp.u-tokyo.ac.jp}

\author{Yukio Hasegawa}
\affiliation{The Institute for Solid State Physics, The University of Tokyo, Kashiwanoha 5-1-5, Kashiwa, Chiba 277-8581, Japan}

\date{\today}% It is always \today, today,

\begin{abstract}
We report on experimental verification of the rotational sense and type of homogeneous chiral spin spiral order in a Mn monolayer on a W(110) substrate using spin-polarized scanning tunneling microscopy. We found that the magnetic contrast due to the spin spiral order almost vanishes with a magnetic tip magnetized normal to the (001) plane, indicating that the spin spiral rotates in the plane. From a shift in the most-contrasted sites by changing the tip magnetization direction within the rotating plane, we reveal that the rotational sense is left-handed, consistent with the previous results predicted by first-principle calculations. By comparing the current system with a chiral magnetic domain wall in Fe double layers on the same substrate, we found that the polarity of the Dzyaloshinskii--Moriya interaction, the driving force of those chiral magnets, is dominantly determined by the choice of the substrate rather than the overlayer.
%\begin{description}
%\end{description}
\end{abstract}

\pacs{Valid PACS appear here}% PACS, the Physics and Astronomy
\maketitle

In magnetic systems whose inversion symmetry is broken, the Dzyaloshinskii--Moriya interaction (DMI) \cite{Dzy57,Moriya60} plays a key role in the formation of chiral spin structures, which include skyrmion lattices \cite{Pfei04,Naga13,Muhl09}, domain walls \cite{Nakatani03, Awano15}, and homogeneous spin spiral structures \cite{Ishi76, Arima06, Togawa12}. Since the inversion symmetry is naturally broken at interfaces, 3d magnetic ultrathin films formed on 5d non-magnetic heavy-elemental substrates, which have strong spin-orbit coupling, often exhibit non-collinear chiral spin structures driven by DMI \cite{Fert90, Heinze11, Meck09, Ryu13, Torr14, Bode07Nat, Ferri08}. These chiral spin structures can exhibit various characteristics depending on the overlayer and substrate materials. They may show various types of rotation, such as Bloch- or N\'eel-type skyrmions \cite{Heinze11}, Bloch or N\'eel domain walls \cite{Meck09,Torr14,Ryu13}, or helical or cycloidal spin spiral structures \cite{Bode07Nat, Ferri08}. They may also have different rotational senses, either left-handed ($\uparrow\leftarrow\downarrow$) or right-handed ($\uparrow\rightarrow\downarrow$). These chiral magnets at interfaces have recently attracted much attention, especially because of their potential application for future spintronics devices \cite{Emori13, Sampaio13, Tomasello14}.

For practical applications, however, it is mandatory to understand how overlayers and substrates affect the polarity and strength of DMI, which determine the characteristics of the chiral magnets. Recent experimental and theoretical studies revealed that the choice of adjacent 5d metal significantly contributes to the characteristics of DMI in the chiral magnets \cite{Torr14,Kashid14}. On the other hand, the contribution of the overlayer material has not been investigated much. Therefore, atomic-scale characterization is desirable for chiral magnets in different overlayers on the same substrate. One of the ideal playgrounds for such studies is the W(110) substrate, on which several chiral magnets were discovered \cite{Bode07Nat, Meck09, Yoshida12, Zimm14}.

Meckler $et$ $al.$ utilized spin-polarized scanning tunneling microscopy (SP-STM) with a triple-axis vector magnet to experimentally determine the rotational sense and type of the domain wall structure in an Fe double layer (DL) on W(110) \cite{Meck09}. However, such detailed experimental verification has not been achieved so far for homogeneous spin spiral structures on the same substrate, such as on a monolayer (ML) and DL of Mn/W(110) \cite{Bode07Nat,Yoshida12} or ML of Cr/W(110) \cite{Zimm14}. 

In this letter, we investigated the rotational sense and type of the homogeneous chiral spin spiral structure in ML Mn/W(110) by SP-STM with a double-axes superconducting magnet. From SP-STM images taken with controlled tip magnetization directions, we reveal that the layer indeed exhibits a cycloidal spin spiral with left-handed rotation, in good agreement with the previous theoretical predictions \cite{Bode07Nat}. By comparison with the rotational sense of Fe DL/W(110), we concluded that the polarity of DMI did not change in these two overlayers formed on W(110).

The experiments were performed in low-temperature ultrahigh vacuum (UHV) STM (Unisoku USM-1300S with RHK R9 controller), in which the sample and the tip were cooled down to 5 K. A two-axis superconducting magnet is equipped to apply magnetic fields perpendicular ($|{\bf B}_\perp| \leq$ 2 T) and parallel ($|{\bf B}_\parallel| \leq$ 1 T) to the sample surface. A W(110) substrate was prepared by several cycles of flashing above 2300 K in UHV and annealing at 1500 K in an oxygen atmosphere of 1$\times$10$^{-4}$ Pa \cite{Bode07surf}. We deposited Mn onto the W(110) substrate for 25 s at a deposition rate of 1.5 ML/min from a Ta crucible heated by electron bombardment. In order to avoid the nucleation of additional layers and to achieve step-flow growth of Mn ML, Mn was deposited just after the flashing to ensure high mobility of the deposited atoms \cite{Bode02PRB}. For the preparation of spin-polarized tips, we deposited Fe by electron bombardment heating on an electrochemically etched W tip that had been flashed in UHV to remove surface oxide layers. The Fe-coated W tips usually exhibit in-plane magnetization at zero magnetic field, and the magnetization can be flipped toward the direction of the external magnetic field with ${\bf B}_\perp$ = $\pm$2 T or ${\bf B}_\parallel$ = $\pm$1 T. \cite{Wie09} All STM and SP-STM measurements were performed in constant current mode.

\begin{figure}
\includegraphics{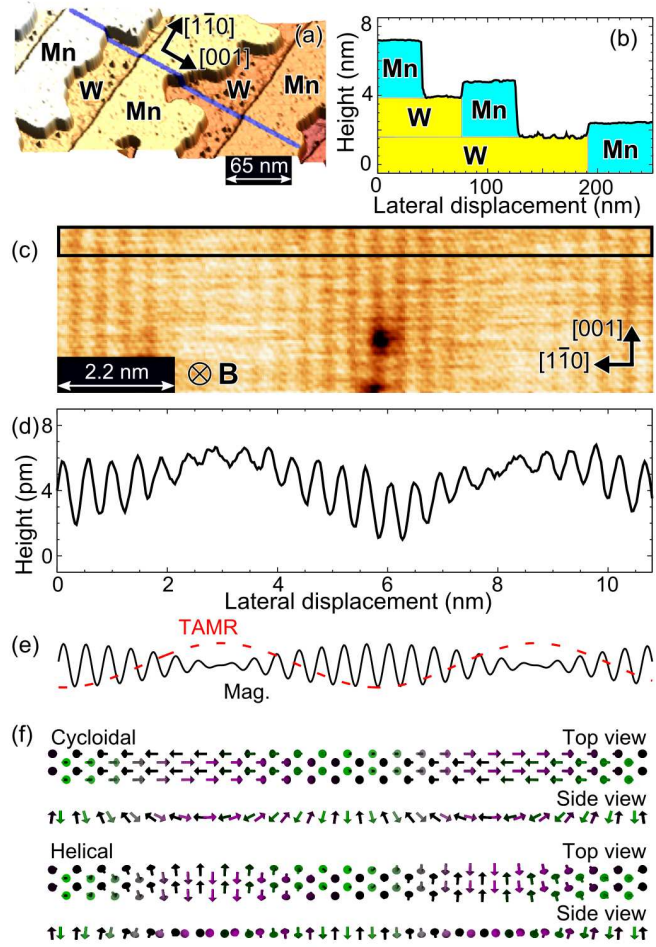}
\caption{\label{fig:epsart}(a) Three-dimensional view of an STM image of a Mn monolayer (ML) formed on a W(110) substrate. (b) Cross-sectional profile taken along the blue line in (a). (c) Spin-polarized STM image of 1 ML Mn/W(110) taken at 5 K with an Fe-coated W tip magnetized perpendicular to the sample surface (sample bias voltage $V_s$ = 10 mV, tunneling current $I_t$ = 10 nA, magnetic field ${\bf B}_\perp$ = 1 T). (d) Cross-sectional profile averaged in the boxed area in (c). (e) Schematics describing decomposition of the profile into tunneling anisotropic magnetoresistance (TAMR, red dashed line) and magnetic (black solid line) contributions. (f) Schematics of two possible spin structures, cycloidal and helical spin spiral. 
}
\end{figure}

\begin{figure}
\includegraphics{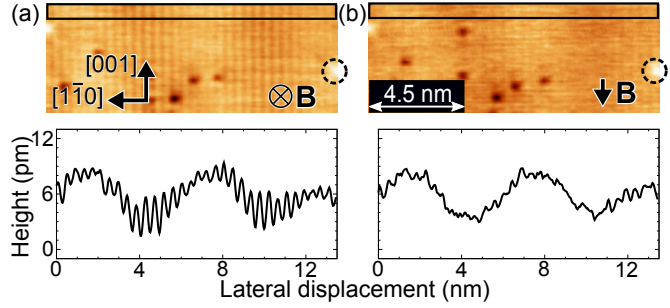}% Here is how to import EPS art
\caption{\label{fig:epsart} SP-STM images of 1 ML Mn/W(110) taken with an Fe-coated W tip sensitive to the (a) out-of-plane and (b) in-plane [001]-oriented sample magnetization component ($V_s$ = 15 mV, $I_t$ = 40 nA,  ${\bf B}_\perp$ = 2 T and  ${\bf B}_\parallel$ = 1 T), and the corresponding cross-sectional profiles averaged in the boxed areas in (a) and (b). Both magnetic and TAMR contrasts are observed in (a), whereas only TAMR contrast is observed in (b). 
}
\end{figure}

\begin{figure*}
\includegraphics{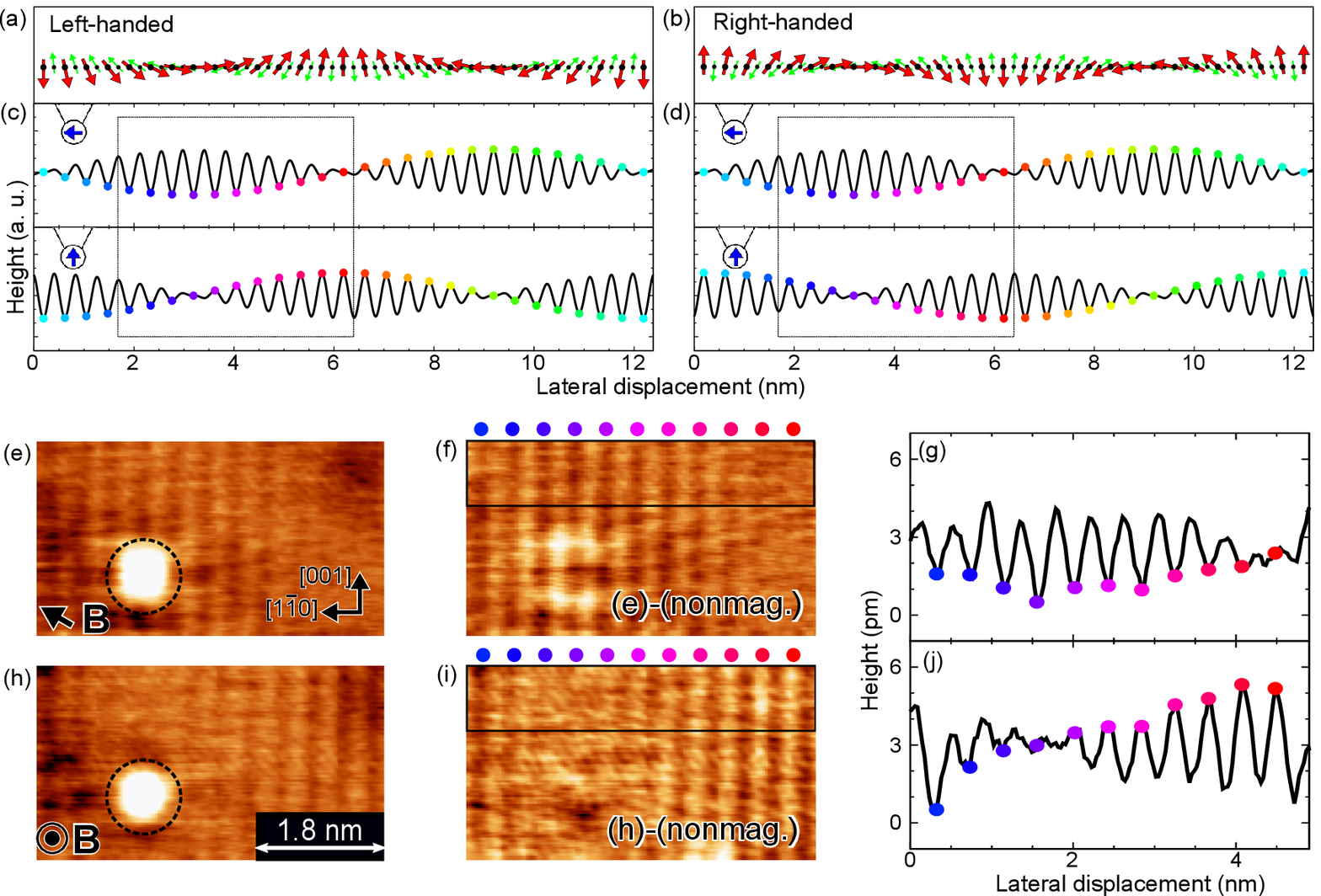}% Here is how to import EPS art
\caption{\label{fig:wide} (a, b) Side views of cycloidal spin spiral structures in two different rotational senses. (c, d) Simulated magnetic contrast profiles of left-handed (c) and right-handed (d) rotating spin spiral structures with two tip magnetization directions. Every two spins are marked with colored dots. (e, h) SP-STM images of 1 ML Mn/W(110) taken in the same area with an Fe-coated W tip magnetized in the in-plane (c) and out-of-plane directions (d) ($V_s$ = 15 mV, $I_t$ = 10 nA, ${\bf B}_\parallel$ = 1 T and ${\bf B}_\perp$ = 2 T). The direction of the applied magnetic field is shown by the arrows. (f, i) Images whose nonmagnetic contribution was subtracted from (e) and (h). The nonmagnetic contribution was derived by averaging images taken at ${\bf B}_\parallel$ = +1 T and ${\bf B}_\parallel$ = -1 T (f) and at ${\bf B}_\perp$ = +2 T and ${\bf B}_\perp$ = -2 T (i). (g, j) Cross-sectional profiles averaged in the boxed areas of (f, i). The same colored dots indicate the same atomic rows in the images of (f, g, i, j).
}
\end{figure*}

A typical STM image taken on 0.65 ML of Mn/W(110) with a non-magnetic W tip is shown in a three-dimensional rendering in Fig. 1(a). Figure 1(b) shows its cross-sectional profile measured along the blue line in Fig. 1(a). Atomically flat terraces of Mn ML grown from the W step edges are observed \cite{Bode02PRB}. Figure 1(c) shows an SP-STM image taken on an ML region of Mn/W(110) with a tip magnetized perpendicular to the sample surface by application of a magnetic field of ${\bf B}_\perp$ = 1 T. The applied magnetic field is strong enough to force the magnetized tip toward the field direction, whereas its influence on the magnetic order is negligible \cite{Bode07Nat, Sessi09}. Since the SP-STM signal depends on the cosine of the angle $\theta$ between the tip and sample magnetization directions, bright and dark rows separated by an atomic distance along the [1\=10] direction indicate that the Mn rows have parallel and antiparallel magnetization components with respect to the tip magnetization direction. The contrast vanishes at the Mn rows whose magnetization directions are close to the in-plane direction (i.e., $\theta$ is close to 90$^\circ$). This indicates a spin-spiral structure as reported in the earlier works \cite{Bode07Nat, Sessi09, Serrate10}. The cross-sectional profile shown in Fig. 1(d), obtained by averaging in the boxed area of Fig. 1(c), presents more detailed features of the contrast. We can decompose the features into two components, a beating pattern and a sinusoidal modulation, as schematically shown in Fig. 1(e). The short (0.45 nm) and long (6 nm) wavelength components in the beating pattern [black solid line in Fig. 1(e)] corresponds to magnetic signals resulting from the antiparallel spin alignment of adjacent [001] Mn rows and the spin spiral structure, respectively. The corrugation is maximized (minimized) in the area where $\theta$ becomes 0$^\circ$ or 180$^\circ$ (90$^\circ$). The sinusoidal modulation (red dashed line in Fig. 1(e)) with a wavelength of 6 nm arises from electronic states induced by the spin-orbit interaction whose density of states depends on the magnetization direction \cite{Berg12}, called tunneling anisotropic magnetoresistance (TAMR) \cite{Bode02PRL, Bode03, Sessi09, Berg12}. The TAMR signal shows the maximum (minimum) at the region where the sample is magnetized to the in-plane (out-of-plane) direction, and it can be observed even with non-magnetized tips. 

Figure 1(f) shows a schematic of two possible spin spiral structures that can reproduce the SP-STM image. The upper panel shows a cycloidal spin spiral structure whose spins are rotating in the (001) plane and the lower one shows a helical spin spiral structure rotating in the (1\=10) plane. In addition, both of them have two possible rotational senses: a left-handed or a right-handed rotation. In the previous work, theoretical calculations based on density functional theory (DFT) suggested that a left-handed cycloidal spin structure is energetically favorable, but the experimental verification has not been done yet \cite{Bode07Nat}. 

To determine the rotational type of the spin structure experimentally, we performed SP-STM measurements with a tip magnetized along the [001] direction. Because of the cosine dependence of the spin-polarized signal on $\theta$ \cite{Wie09}, the intensity of the magnetic contrast in this measurement should be considerably suppressed for the cycloidal structure but remain the same for the helical one. Therefore, this measurement provides conclusive evidence to determine the type of spin spiral structure. The upper panels of Figs. 2(a) and 2(b) show SP-STM images obtained with tips sensitive to the out-of-plane direction and along the [001] direction, respectively. We obtained the two images in the same field of view, as evidenced by a white-dot marker (circled with a dashed line) that appears on the right edge of the images. Some of the impurities (i.e., dark spots) changed their positions during the imaging. Both the magnetic and TAMR components are visible with the out-of-plane tip [Fig. 2(a)]. In contrast, in the SP-STM image taken with the [001]-magnetized tip [Fig. 2(b)], the magnetic contrast is strongly suppressed, whereas the TAMR signal stays the same, as more clearly demonstrated in the averaged cross-sectional profiles (the lower panels) collected in the boxed areas of Figs. 2(a, b). A faint magnetic contrast found in Fig. 2(b) is presumably due to small misalignment of the tip magnetization direction. Based on these observations, we conclude that Mn ML has a cycloidal spin spiral structure, a conclusion that is consistent with the DFT calculations \cite{Bode07Nat}. This result is also consistent with the mechanism of the interfacial DMI generated between a heavy-elemental substrate and a 3d magnetic layer, which only produces chiral spin structures rotating in the plane normal to the interface and parallel to the propagating axis, e.g., cycloidal spin spiral or N\'eel type domain walls \cite{Fert13}.

Then, in order to clarify the rotational sense of the cycloidal spin spiral structure, we performed SP-STM measurements with spin-polarized tips magnetized in two orthogonal directions within the (001) plane. 
We show the side views of two possible spin spiral structures, left- and right-handed ones, in Figs. 3(a, b), respectively, and the corresponding simulated SP-STM signals depending on the tip magnetization direction in Figs. 3(c, d), respectively.
The colored dots are located at every two adjacent spins in Figs. 3(c, d) to identify the same atomic spins in the profiles. In the spin spiral structures, the rotational angle between the nearest neighbor (next-nearest neighbor) spins along the [1\={1}0] axis is +173$^\circ$ (-14$^\circ$). When we focus on every two adjacent spins, therefore, the magnetic contrast exhibits a sinusoidal variation, as can be seen by following the colored dots. By changing the direction of the tip magnetization from parallel to perpendicular to the surface in the (001) plane, [i.e., from the leftward to the upward direction, as shown with arrows in Figs. 3(c, d)], we should observe +90$^\circ$ or -90$^\circ$ phase shifts of the sinusoidal profile depending on the rotational sense. As shown in Figs. 3(c, d), if the rotational sense is left-handed (right-handed), the sinusoidal profile should shift to the left (right). We can therefore determine the rotational sense through observation of such a phase shift.

Figures 3(e) and (h) show SP-STM images taken in the same area with the tips magnetized in the directions shown in the figures. White clusters, marked with dotted circles in Figs. 3(e, h), provide proof that it is the same area. A phase shift in the beating pattern is observed between the two images. In order to extract the magnetic contrast, we subtracted the nonmagnetic contributions (i.e., structural and TAMR components) from the two images, as shown in Figs. 3(f, i). The nonmagnetic contributions are derived by averaging two SP-STM images taken in the same area with the tips magnetized in opposite directions \cite{Kube05}. For instance, the magnetic contrast shown in Fig. 3(i) is extracted by calculating a subtraction of the average of two SP-STM images taken at ${\bf B}_\perp$ = +2 T and -2 T from the image shown in Fig. 3(h). The marked white clusters disappear in the averaged images because the structural contrast does not depend on the tip magnetization direction. Cross-sectional profiles taken in the boxed areas of Figs. 3(f, i) are shown in Figs. 3(g, j). We placed the colored dots in the same manner as in Figs. 3(c, d), and the same part of the sinusoidal profiles as those of Figs. 3 (g, j) are marked with boxes in Figs. 3(c, d). The direction of the phase shift shown in Figs. 3(g, j) is consistent with that of Fig. 3(c), indicating that the rotational sense of the spin spiral structure is left-handed. This result is consistent with the prediction by DFT calculations \cite{Bode07Nat}.

As mentioned before, the domain walls in the Fe DL/W(110) were determined as N\'eel-type domain walls with right-handed rotation propagating along to the [001] axis \cite{Meck09}. These two results observed on W(110), left-handed for Mn ML and right-handed for Fe DL, on first glance, seem contradictory to the previous studies, which found that the polarity of DMI was dominantly determined by the substrate \cite{Torr14,Kashid14}.
To understand the situation, we carefully considered the competition between the exchange interaction and DMI in both systems. DMI can produce two types of perpendicular spin orientations depending on the polarity of DMI, i.e., right-handed DMI ($\uparrow\rightarrow$) or left-handed DMI ($\uparrow\leftarrow$).
We considered that the left-handed rotation of Mn ML ($\uparrow\searrow\leftarrow\nearrow\downarrow$) is a consequence of the competition between antiferromagnetic interaction ($\uparrow\downarrow$) and right-handed DMI ($\uparrow\rightarrow$). The right-handed rotation of Fe DL ($\uparrow\nearrow\rightarrow\searrow\downarrow$) can be considered the consequence of the competition between ferromagnetic interaction ($\uparrow\uparrow$) and right-handed DMI ($\uparrow\rightarrow$). Therefore, we finally conclude that the consistent role of the substrate on the polarity of DMI is also true for these systems on W(110), where the contributions of the overlayers seem insignificant.

In conclusion, we determined the rotational type and sense of the spin spiral structure observed on Mn ML/W(110) experimentally using SP-STM with an Fe-coated magnetic tip in magnetic fields parallel and perpendicular to the sample surface. With the tip magnetized along the (001) plane, we found strong suppression of the magnetic contrast, indicating that the spin structure is cycloidal spin spiral rotating in the (001) plane. We also revealed the left-handed rotation by observing the phase shift in the magnetic contrast of the SP-STM images taken with spin-polarized tips magnetized in various directions in the (001) plane. Our experimental results agree well with the spin spiral structure predicted by DFT calculations. 
Moreover, we revealed that the polarity of DMI is same for both Mn ML and Fe DL on W(110).
This suggests that the substrate plays the dominant role in determining the polarity of DMI for these systems, but the contributions of the overlayers seem insignificant. This work also demonstrates that SP-STM measurements with a two-axis superconducting magnet enable us to explicitly determine complicated surface spin structures.

\begin{acknowledgments}
We thank Takahisa Arima, Koji Hukushima, Howon Kim and Kohji Nakamura for fruitful discussion. We thank Editage (www.editage.jp) for English language editing. This work was partially supported by JSPS KAKENHI Grants No. 25707025, No. 26110507, and No. 26120508 (Grants-in-Aid for Scientific Research on Innovative Areas "Initiative for high-dimensional data-driven science through deepening of sparse modelling").
\end{acknowledgments}

\bibliography{apssamp.bib}% Produces the bibliography via BibTeX.

\end{document}